\documentclass[aps,prb,twocolumn,showpacs,superscriptaddress]{revtex4}
\usepackage{graphicx}

\begin{document}

\title {Incommensurate magnetic structure in the orthorhombic
perovskite ErMnO$_3$}

\author{F. Ye}
\email{yef1@ornl.gov}
\affiliation{Neutron Scattering Science Division,
Oak Ridge National Laboratory, Oak Ridge, Tennessee 37831-6393 }

\author{B.~Lorenz}
\affiliation{Department of Physics and TCSUH, University of
Houston, Houston, Texas 77204-5002, USA}

\author{Q.~Huang}
\affiliation{NIST Center for Neutron Research, National Institute
of Standards and Technology, Gaithersburg, Maryland 20899}

\author{Y.~Q.~Wang}
\affiliation{Department of Physics and TCSUH, University of
Houston, Houston, Texas 77204-5002, USA}

\author{Y.~Y.~Sun}
\affiliation{Department of Physics and TCSUH, University of
Houston, Houston, Texas 77204-5002, USA}

\author{C.~W.~Chu}
\affiliation{Department of Physics and TCSUH, University of
Houston, Houston, Texas 77204-5002, USA}

\author{J.~A.~Fernandez-Baca}
\affiliation{Neutron Scattering Science Division,
Oak Ridge National Laboratory, Oak Ridge, Tennessee 37831-6393 }

\author{Pengcheng Dai}
\affiliation{Department of Physics and Astronomy,
The University of Tennessee, Knoxville, Tennessee 37996-1200}
\affiliation{Neutron Scattering Science Division,
Oak Ridge National Laboratory, Oak Ridge, Tennessee 37831-6393 }

\author{H.~A.~Mook}
\affiliation{Neutron Scattering Science Division,
Oak Ridge National Laboratory, Oak Ridge, Tennessee 37831-6393 }

\date{\today}

\begin{abstract}
By combining dielectric, specific heat, and magnetization
measurements and high-resolution neutron powder diffraction, we have
investigated the thermodynamic and magnetic/structural properties of
the metastable orthorhombic perovskite ErMnO$_3$ prepared by high-pressure 
synthesis. The system becomes antiferromagnetically
correlated below 42~K and undergoes a lock-in transition at 28~K
with propagation wave vector $(0,k_b,0)$, which remains
incommensurate at low temperature. The intercorrelation between the
magnetic structure and electric properties and the role of the rare
earth moment are discussed.
\end{abstract}

%

\maketitle

Multiferroic materials (in which magnetism and ferroelectricity
coexist)
materials have attracted great attention in recent years because of
their fundamental as well as technical
importance.\cite{kimura03,hur04,lawes05,ikeda05,kimura06,taniguchi06}
Among these, the frustrated magnets are considered the most
promising candidates to achieve mutual control of
magnetic and ferroelectric properties. For example, the rare-earth
perovskite manganites RMnO$_3$ (R=Tb, Dy) exhibit flopping of the
electric polarization ($P$) with applied magnetic field $H$, or
reversal of the magnetic helicity by electric field
$E$.\cite{kimura03,yamasaki07} Despite the continuing discovery of
new multiferroic materials, the understanding of the fundamental
mechanism of the magnetic-ferroelectric coupling is far from
completed and remains a
challenge.\cite{khomskii06,eerenstein06,cheong07} One possible
microscopic mechanism of ferroelectricity (FE) induced by the
complex magnetic order involves a noncollinear coupling between a
uniform electric polarization $P$ to an inhomogeneous magnetic order
$M$. The symmetry-allowed term $PM\partial M$ breaks the space
inversion symmetry and gives rise to electric polarization as soon
as magnetic ordering of a proper kind sets in. Generally speaking, a
spiral or helical magnetic structure in frustrated magnets with
appropriate crystal symmetry leads to an electric polar
state.\cite{katsura05,sergienko06prb,mostovoy06}

However, the spiral magnetic structure is not the only source 
of FE. It was proposed that a collinear $E$-type magnetic
structure in the orthorhombic perovskite manganites and nickelates
allows a finite ferroelectric
polarization.\cite{sergienko06} The estimated $P$ might be orders
of magnitude larger than that in the helical magnets, because the
underlying mechanism does not involve spin-orbit interactions or
noncollinear spin structures. Instead, it is the competition between
the elastic energy and the energy gain due to the virtual hopping of
$e_g$ electrons within the zigzag spin chains that causes the
coherent displacement of the oxygen atoms in the ground state.
Indeed, giant magnetoelectric response and spontaneous electric
polarization have been reported in the orthorhombic HoMnO$_3$ and
YMnO$_3$ manganites, in which the FE is established simultaneously
with the collinear magnetic structure.\cite{lorenz04,lorenz06}
Furthermore, there is considerable enhancement of electric
polarization $P$ in HoMnO$_3$ when the holmium ions order below
15~K, strongly suggesting the involvement of the rare-earth
moment.\cite{lorenz06} To extend our search for possible FE
materials as well as to assess the role of the rare-earth moment in
stabilizing the FE, we have prepared orthorhombic perovskite
ErMnO$_3$ manganite where the Er$^{3+}$ moment orders at much lower
temperature. We find that the low-$T$ magnetic structure of
orthorhombic ErMnO$_3$ is incommensurate (ICM), in contrast to the
generally accepted $E$-type phase for this distorted
manganite.\cite{zhou06,tachibana07} Antiferromagnetic
fluctuation sets in at $T_{N1}\approx 42$~K with wave vector of
$(0,k_b,0)$. The wave vector locks onto a fixed value of
$q_m=(0,0.433,0)$ below $T_{N2}\approx 28$~K. Even at the lowest
temperature, the magnetic order remains short range.  Consistent
with this, the FE in ErMnO$_3$ is relatively weak compared to
HoMnO$_3$ and YMnO$_3$. This highlights the close connection between
the emergence of ferroelectricity and the long-range magnetic
structure.

The powder specimens of $\rm ErMnO_3$ were prepared using solid-state
reaction methods.\cite{lorenz04} The hexagonal compounds were
transformed into the orthorhombic structure by high-pressure
sintering for 5 h (1100 $^\circ$C, 3.5 GPa). The specific heat and
dielectric properties were measured employing physical property
measurement system (Quantum Design) for temperature control, using
the heat capacity option as well as a home-made capacitance probe
adapted to the cryostat, respectively. For dielectric measurements,
the sample was shaped as a parallel plate capacitor and silver paint
was used as electrodes. Magnetic properties were measured in a
magnetic property measurement system (Quantum Design). Neutron
powder diffraction (NPD) patterns were collected on the
high-resolution, 32-counter BT-1 diffractometer and BT-7 triple-axis
spectrometer at the NIST Center for Neutron Research (NCNR). A total
mass of 1.5 gram was used for the NPD measurement.

\begin{figure}[ht!]
\includegraphics[width=3.2in]{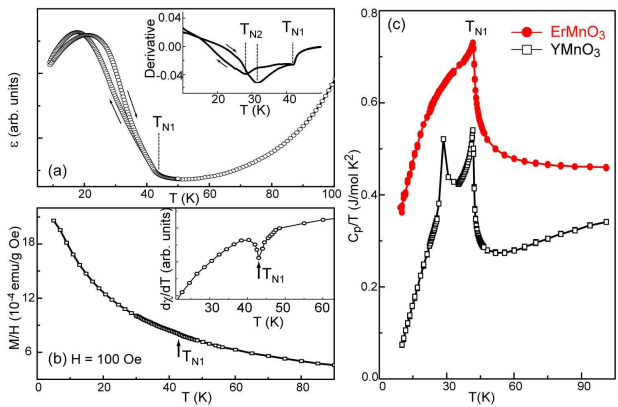}
\caption{\label{fig:bulk} (Color online) Temperature dependence of
(a) the dielectric constant $\varepsilon(T)$, (b) the magnetization
$M(T)$, and (c) the specific heat $C_p(T)$ of the orthorhombic
ErMnO$_3$.  Inset shows the $T$ derivative of $\varepsilon(T)$ in
the lower temperature range. $C_p(T)$ data for the orthorhombic YMnO$_3$
(similar to the work reported by Kim {\it et al.},
\onlinecite{kim06}) are also plotted for comparison. 
}
\end{figure}

Figures 1(a)-1(c) display the temperature dependence of the
dielectric constant $\varepsilon(T)$, magnetization $M(T)$, and the
specific heat $C_p(T)$ of the orthorhombic ErMnO$_3$. Below the
N\'{e}el temperature, $\varepsilon(T)$ increases rapidly and passes
through a broad maximum at low $T$. Two transitions are clearly seen
in the temperature derivative $d\varepsilon(T)/dT$, where $T_{N1}$
is well defined at 42~K and a second transition takes place near
30~K with strong hysteresis, indicating its first-order character.
Similarly, $M(T)$ shows a small anomaly at $T_{N1}$, while that
transition can be better seen in $d\chi/dT$. The signals of $M(T)$
at lower temperatures are dominated by the paramagnetic moment from
the rare-earth erbium ions. Figure~1(c) compares the specific heat
$C_p(T)$ of ErMnO$_3$ with that of the orthorhombic YMnO$_3$ in
which spontaneous electric polarizations occurs.\cite{lorenz06}
While $C_p(T)$ of YMnO$_3$ shows two sharp peaks which correspond to
the onset of antiferromagnetic (AFM) correlations and a lock-in
transition to long-range magnetic order with the onset of
ferroelectricity,\cite{munoz02,kim06,lorenz06} the $C_p(T)$ of ErMnO$_3$
only displays a sharp peak near 42~K; the second transition near
$T_{N2}$ is less discernible.

\begin{figure}[ht!]
\includegraphics[width=3.2in]{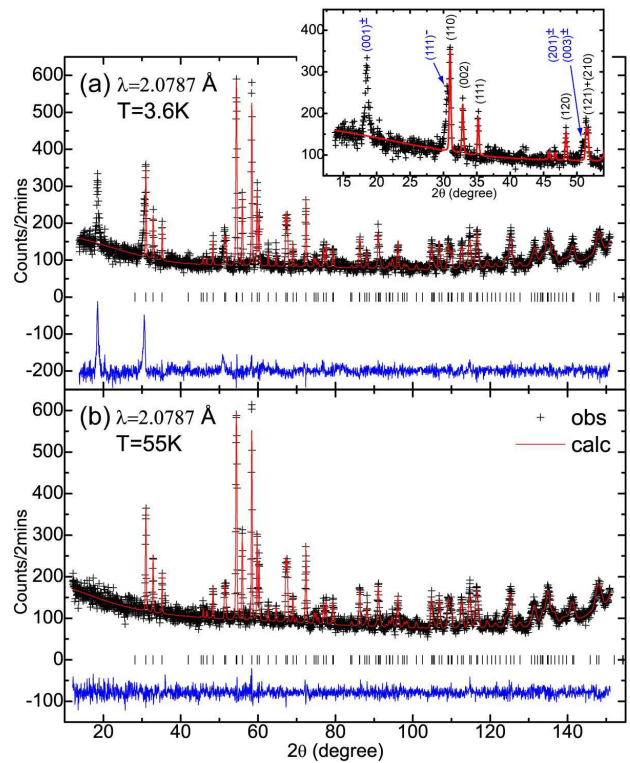}
\caption{\label{fig:NPD} (Color online) (a) High-resolution neutron
powder diffraction pattern of the orthorhombic ErMnO$_3$
specimen at (a) 3.6~K (below $\rm T_{N1}$) and (b) 55~K (above $\rm
T_{N1}$). Inset zooms in on the lower-angle portion of the diffraction
pattern, with the magnetic and structural reflections indexed. 
}
\end{figure}

To determine the magnetic structure and its correlation with the
bulk measurement results, we performed NPD at a series of
temperatures near the transition and at room temperature with incident
neutron wavelength $\lambda=2.0787~\AA$. The nuclear structure peaks
can be described by a single phase with space group $Pbnm$ at all
temperatures. No hexagonal phase could be detected  within the
experimental error. Figures 2(a) and 2(b) plot the representative
diffraction profiles at 3.6 and 55~K.  As revealed by the powder
refinement (Table~I), the smaller lattice parameters and Mn-O-Mn
bonding angles are consistent with the evolution of ionic size of
the rare-earth atoms.\cite{alonso00,zhou06} Below 42~K, additional
peaks appear at $2\theta$ angles which are not allowed in the $Pbnm$
symmetry.  This reveals the appearance of ICM magnetic
fluctuations consistent with the anomalies present in
$\varepsilon(T)$, $M(T)$, and $C_p(T)$. As temperature is further
lowered, the corresponding magnetic peaks move slightly. At 3.6~K,
strong magnetic peaks become evident at several positions [inset of
Fig.~2(a)]. We find that the magnetic modulation is along the
crystalline $b$ axis with the propagation wave vector
$q_m=(0,k_b,0)$. The magnetic satellite peaks occur near the nuclear
Bragg peaks $(h,k,l)$ with observed reflection conditions of
$h+k=2n$ and $l=2n+1$.  The indexing of the magnetic peaks agrees
with previous neutron diffraction results of the orthorhombic
YMnO$_3$.\cite{quezel74,munoz02}  The peaks located at
18.60$^\circ$, 30.67$^\circ$, and 51.04$^\circ$ correspond to the
reflections from $(0,\pm k_b,1)$, $(1,1-k_b,1)$, $(0,\pm k_b,3)$, and
$(2,\pm k_b,1)$.

\begin{table}[ht!]
    \caption{Structural parameters of orthorhombic ErMnO$_3$ from
    Rietveld refinement at 3.6, 55, and 300~K. Atomic
    positions [Er $4b(x,y,0.25)$, Mn $4c(0.5,0,0)$, O$_1$
    $4c(x,y,0.25)$, O$_2$ $8d(x,y,z)$]. $R_p$ is the residual, and
    $R_{wp}$ is the weighted residual. The temperature factor
    $U_{iso}$ of Mn is fixed in the refinement.
    }
    \label{tab1}
\begin{tabular}{llrrr}
\hline \hline
      &  & 3.6~K$\quad$   	& 55~K$\quad$  	& 300~K       	\\
S. G. &      & $Pbnm$$\quad$    & $Pbnm$$\quad$	&$Pbnm$ 	\\
    \hline
$\qquad a(\AA)$   	&     	&5.2273(2)$\quad$ &5.2270(2)$\quad$ &5.2315(3)\\
$\qquad b(\AA)$   	&     	&5.7922(2)$\quad$ &5.7926(2)$\quad$ &5.7987(4)\\
$\qquad c(\AA)$   	&     	&7.3277(3)$\quad$ &7.3308(3)$\quad$ &7.3410(4)\\
$\qquad V(\AA^3)$ 	&     	&221.865(12)$\quad$ &221.960(11)$\quad$ &222.699(31)\\
Mn\ $\, U_{iso}(\AA^2)$&&0.0030$\quad$		&0.0030$\quad$ 	&0.0030\\
Er\ $\;\,\, x$ 	&	&-0.0171(9)$\quad$	&-0.0174(9)$\quad$	&-0.0188(10)\\
$\qquad y$  	&	& 0.0840(7)$\quad$    	& 0.0846(7)$\quad$ 	&0.0844(7)\\
$\qquad U_{iso}(\AA^2)$&&0.0038(14)$\quad$	&0.0044(13)$\quad$ 	&0.0043(10)\\
O$_1$\ $\;\, x$	&     	& 0.113(1)$\quad$   	& 0.113(1)$\quad$	&0.113(1)\\
$\qquad y$  	&     	& 0.461(1)$\quad$   	& 0.462(1)$\quad$	&0.462(1)\\
$\qquad U_{iso}(\AA^2)$&&0.0019(12)$\quad$	& 0.0018(11)$\quad$ 	&0.0058(8)\\
O$_2$\ $\;\, x$	&     	& 0.6984(8)$\quad$    	& 0.6989(7)$\quad$	&0.7000(8)\\
$\qquad y$ 	&    	& 0.3268(7)$\quad$    	& 0.3278(7)$\quad$	&0.3266(8)\\
$\qquad z$ 	&     	& 0.0541(6)$\quad$    	& 0.0555(6)$\quad$	&0.0536(6)\\
$\qquad U_{iso}(\AA^2)$&& 0.0019(12)$\quad$	& 0.0018(11)$\quad$ 	&0.0058(8)\\
  Mn-O$_1$-Mn(deg) &   	& 141.2(4)$\quad$     	&142.1(3)$\quad$ 	&142.0(3)\\
  Mn-O$_2$-Mn(deg) &   	& 143.7(3)$\quad$     	&143.2(2)$\quad$ 	&144.0(2)\\
$Rp(Rwp)(\%)$	&    	&7.13(8.90)$\quad$  	&6.55(8.27)$\quad$ &6.70(8.39)\\
    $\chi^2$    &    	&0.91$\quad$        	&0.81$\quad$       	&0.82\\
    \hline \hline
    \end{tabular}
\end{table}

Figures 3(a)-3(c) summarize the thermal evolution of the structural
and magnetic properties derived from the NPD analysis.  The crystal
structure remains orthorhombic throughout the entire measured
temperature range [Fig.~3(a)]. The lattice parameters and cell
volume show smooth variations.  The typical lattice distortion
across the transition due to the magnetoelastic coupling observed in
many magnetically frustrated system is less
evident.\cite{chapon04,ye06} However, a sudden collapse in bonding
angle of Mn-O$_1$-Mn (along the $c$ axis) near $T_{N1}$ does
indicate the onset of AF order with 180$^\circ$ phase flipping in
that direction, while the in-plane Mn-O$_2$-Mn angle shows no sign of
AF correlation near $T_{N1}$.  Figure~3(b) displays the order
parameter measured by monitoring the peak intensity at
$2\theta=18.6^{\circ}$, the corresponding wave vector of the
strongest magnetic peaks at $(0,\pm k_b,1)$.  The intensity
increases below $T_{N1}$ with little hysteresis upon cooling and
warming. Figure~3(c) plots the $T$ dependence of the magnetic
modulation. The wave vector occurs at $(0,k_b,0)$ with $k_b \approx
0.415$ just below $T_{N1}$, and gradually shifts to a larger value with
decreasing temperature. It locks onto $k_b=0.433$ at $T_{N2}=$28~K
and remains constant at low temperature. The evolution of
the magnetic modulation bears a close resemblance to that of the
orthorhombic YMnO$_3$,\cite{munoz02} in which the ICM magnetic order
locks onto a wave vector with $k_b=0.435$ below 28~K.\cite{icm}
Although the magnetic phase transition of ErMnO$_3$ is manifested by
the sharp anomaly near $T_{N1}$ in the thermodynamics measurements,
the magnetic correlation length remains finite even at the lowest
temperature. As shown in Fig.~3(d), the magnetic peaks at
$(0,\pm k_b,1)$ and higher scattering angles have Lorentzian
profiles indicative of short-range order.  The magnetic
correlation length $\xi$ is estimated using the method described in
Ref.~\onlinecite{correlation}:
\begin{equation}
\xi=\pi/\sqrt{2\ln{2}}\ \sigma(q_0)
\end{equation}
where $\sigma(q_0)$ is the intrinsic width of the magnetic peak with
$q_0$ in units of $\AA^{-1}$. We obtain $\xi \approx 210~\AA$, much
shorter than the instrumental resolution of $1000~\AA$. 

\begin{figure}[ht!]
\includegraphics[width=3.2in]{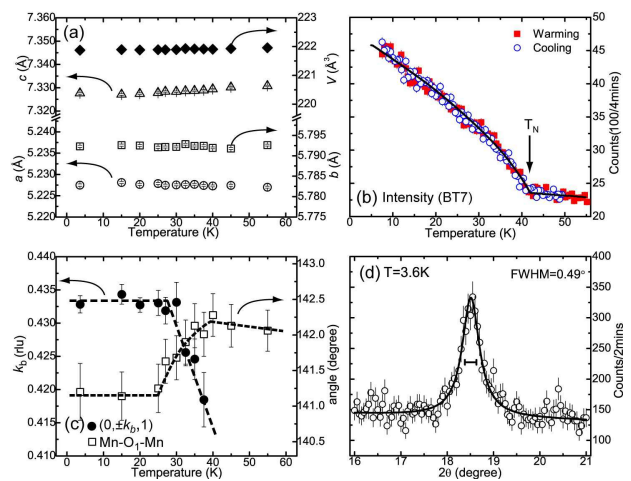}
\caption{\label{fig:Tdep}
(Color online) Temperature dependence of (a) the lattice parameters
$a$, $b$, $c$, and cell volume $V$, (b) the peak intensity of
$q=(0,\pm k_b,1)$, and (c) the $k_b$ component of the propagation
wave vector and the Mn-O$_1$-Mn bonding angle. (d) The scattering
profile near $q=(0,\pm k_b,1)$ at $T$=3.6~K.
}
\end{figure}

It was initially proposed that YMnO$_3$ forms a helical spin
structure with Mn moments within the $ac$ plane and the rotation
axis along the $b$ direction.\cite{quezel74} Later neutron
diffraction measurements suggest that the Mn spins are sinusoidally
modulated, with both the moment direction and modulation wave vector
along the $b$ axis in the $Pbnm$ space group.\cite{munoz02} For
ErMnO$_3$, it is difficult to definitively determine the magnetic
structure, given the short-range nature of the magnetic order and
limited number of magnetic peaks. Nevertheless, the similar
propagation wave vector and $T$ dependence of the modulation
wave vector suggest that both ErMnO$_3$ and YMnO$_3$ have the same
magnetic structure. 

It is interesting to compare the magnetic and ferroelectric
properties of three rare-earth perovskite manganites RMnO$_3$ (R=Ho,
Y, and Er). Neutron diffraction measurements show that HoMnO$_3$
possesses an E-type magnetic structure\cite{brinks01,munoz00} which
is verified theoretically to be stable at low
temperature.\cite{picozzi06} The fact that both HoMnO$_3$ and
YMnO$_3$ exhibit spontaneous electric polarization at the magnetic
lock-in transitions into either commensurate E-type or ICM magnetic
order strongly indicates a close interplay between the magnetism
and ferroelectricity. According to Sergienko {\it et
al.}, the cooperative movement of oxygen atoms resulting from the
competition between elastic energy (favoring a distorted
Mn-O-Mn bonding angle) and energy gain by the virtual hopping of
$e_g$ electrons within the ferromagnetic chain (favoring 180$^\circ$
bonding angle) leads to polarization $P$ perpendicular to
the magnetic modulations. In YMnO$_3$, the virtual hopping mechanism
may still give rise to a local polarization within one ferromagnetic
chain but due to the ICM modulation along the $b$ axis this
polarization will be modulated and reverse sign with the magnetic
modulation. The origin of the observed macroscopic polarization in
YMnO$_3$ is not yet explained and still a matter of discussion.
In the case of ErMnO$_3$, the lack of long-range magnetic
correlation (with magnetic domain size around 200~$\AA$)
significantly affects the ferroelectric properties. Our preliminary
measurements of the pyroelectric current have not been able to prove the
existence of a spontaneous polarization above the resolution of the
measurement. The weakness of macroscopic polarization is likely
linked to the absence of long-range magnetic order. 
One possible explanation of the magnetic fluctuations is the
enhanced magnetic frustration in ErMnO$_3$. We notice considerable
increase of Jahn-Teller distortion in the perovskite RMnO$_3$ with
small R ionic size.\cite{tachibana07} The in-plane bonding angle of
Mn-O$_2$-Mn decreases from 144.1$^\circ$ (HoMnO$_3$) and
144.6$^\circ$ (YMnO$_3$) to 143.9$^\circ$ for
ErMnO$_3$.\cite{alonso00} Such distortion effectively increases the
magnetic frustration due to the competition between the 
nearest-neighbor FM interaction and the next-nearest-neighbor AFM
interaction in the $ab$ plane.\cite{kimura03prb} The short-range
magnetic order of Mn spins could also be related to the paramagnetic
Er$^{3+}$ ions in that temperature range. These paramagnetic
fluctuations might not be sufficient to support the long-range
magnetic order at the Mn sites.  The contribution of the rare-earth
magnetic moment is further emphasized in HoMnO$_3$. The strong
external field dependence of the ferroelectric polarization in that
material reveals the active role of the holmium moment order in
stabilizing the magnetic structure and enhancing of the FE. A
Similar feature has been reported in the multiferroic perovskite
DyMnO$_3$,\cite{prokhnenko07} where the ICM order of Dy moments
closely tracks the evolution of the FE polarization and suppresses
it when the Dy moment becomes commensurate. 

In summary, we have used dielectric, magnetization, and specific heat
measurements as well as high-resolution neutron powder diffraction
to investigate the thermodynamic and magnetic properties of the
orthorhombic perovskite ErMnO$_3$. The system forms ICM AF
correlation below 42~K and undergoes a lock-in transition at 28~K with
propagation wave vector of (0,0.433,0). The magnetic fluctuations
remain short range at low $T$. The correlation of the magnetic and
ferroelectric properties is also discussed.

We thank I. Sergienko for useful discussions. Oak Ridge National
Laboratory is managed by UT-Battelle, LLC, for the US Department of
Energy under contract No.~DE-AC05-00OR22725. The work is supported
by the US DOE BES under contract No.~DE-FG02-05ER46202, the T.L.L.
Temple Foundation, the J. J. and R. Moo res Endowment, and the State
of Texas through TCSUH.


\begin{references}
\bibitem{icm} An accurate determination of the propagation
wave vector $(0,k_b,0)$ is difficult even below the ``lock-in''
transition. However, the similar magnetic behavior between ErMnO$_3$
and YMnO$_3$ indicates that the magnetic structure of ErMnO$_3$ at
low temperature is also incommensurate.
\end{references}
\end{document}